\documentclass[11pt]{article}

\usepackage{cite}
\usepackage{epsf}
\usepackage{amsmath}
\usepackage{amsfonts}
\usepackage{amssymb}

\usepackage{graphicx}
\usepackage{epsfig}
\usepackage{latexsym}

\def\bbox#1{\mbox{\boldmath$#1$}}

\def\corresponds{{\lower.2ex\hbox{=}}{\rm\kern-.75em^\triangle}}
\def\succsim{\succ\kern-.9em_\sim\kern.3em}
\def\precsim{\prec\kern-1em_\sim\kern.3em}
\def\slantfrac#1#2{\kern1em^{#1}\kern-.3em/\kern-.1em_{#2}}
\def\lfrac#1#2{{}^{#1\!}\kern-.0em/_{#2}}

\def\buildrel#1\under#2{\mathrel{\mathop{\kern0pt #2}\limits_{#1}}}

\begin{document}

\vspace*{1.0cm}
\begin{center}
\begin{tabular}{c}
\hline
\rule[-5mm]{0mm}{15mm}
{\Large \sf Two--Loop Bound--State Calculations}\\
{\Large \sf and Squared Decay Rates}\\[2ex]
\hline
\end{tabular}
\end{center}
\vspace{0.2cm}
\begin{center}
Ulrich D. Jentschura$^{a)}$, Christoph H. Keitel$^{a)}$ 
and Krzysztof Pachucki$^{b)}$\\[1ex]
{\it $^{a)}$Fakult\"{a}t Physik der Albert--Ludwigs--Universit\"{a}t,\\
Theoretische Quantendynamik,\\
Hermann--Herder--Stra\ss{}e 3, D--79104 Freiburg, Germany}\\[1ex]
{\it $^{b)}$Institute of Theoretical Physics,\\
University of Warsaw,
ul.~Ho\.{z}a 69, 00-681 Warsaw, Poland}
\end{center}
\vspace{0.3cm}
\begin{center}
\begin{minipage}{11.8cm}
{\underline{Abstract}}
We discuss the $\epsilon$-method as used in various recent
QED bound-state calculations by considering
mathematical model examples. Recently obtained results for
higher-order self-energy binding corrections at the
two-loop level are reviewed.
Problems associated with the interpretation of
squared decay rates as radiative bound-state energy level shifts
are discussed. We briefly expand on the relation
of squared decay rates to nonresonant and radiative
corrections to the Lorentzian line shape, including their
dependence on the experimental process under study.
\end{minipage}
\end{center}
\vspace{1.3cm}

quantum electrodynamics -- specific calculations.

\newpage

%
%
\section{Introduction}
\label{Introduction}

\begin{figure}[htb]
\begin{center}
\begin{minipage}{14cm}
\centerline{\mbox{\epsfysize=2.5cm\epsffile{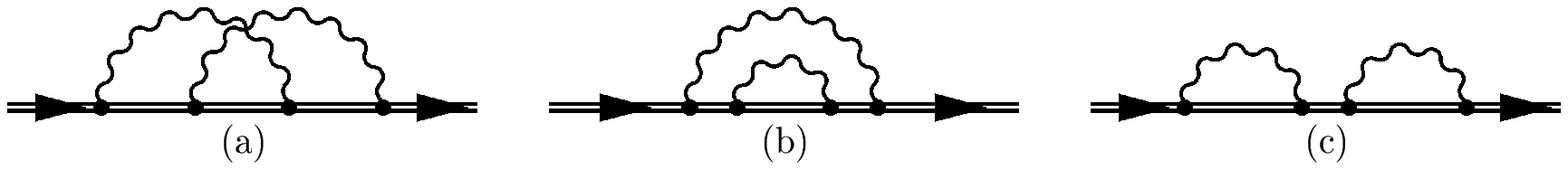}}}
\caption{\label{fig1} 
The crossed (a), rainbow (b) and the
loop-after-loop diagram (c) which contribute to the two-loop
self-energy for a bound electron. The propagator of the bound
electron is denoted by a double line.}
\end{minipage}
\end{center}
\end{figure}

This paper is concerned with mathematical methods employed
in recent analytic evaluations~\cite{Pa1993,JePa1996,JeSoMo1997,
Pa2001,JePa2002,JeNa2002} of higher-order binding corrections
to the Lamb shift. These methods rely on a separation of the 
virtual photon energy integration into high- and low-energy
domains. The methods are applicable in a wider context, and
we attempt to provide a certain clarification by considering
mathematical model examples. We focus on the two-loop self-energy
correction (see Fig.~\ref{fig1}) in hydrogenlike systems
with a low nuclear charge number. We also discuss related corrections
in helium.

In the second part of the paper, we discuss a recent 
investigation~\cite{JeEvKePa2002} which is concerned with 
predictive limits of energy shifts as derived from the 
Gell--Mann--Low--Sucher Theorem~\cite{GMLo1951,Su1957}. 
Expressions obtained based on this theorem have provided the 
basis of level-shift calculations for decades; these may not
be accurate enough for projected future experiments. 
Certain problems associated with this theorem find a rather natural
solution in the two-time Green function method~\cite{Sh2002},
other problematic aspects connected with this theorem concern
the interpretation of level shifts involving squared decay rates.

Our investigations are motivated by the recent
dramatic progress in laser-spectroscopic experiments
in atomic hydrogen (e.g.~\cite{BeEtAl1997,ScEtAl1999,NiEtAl2000}) 
as well as helium~\cite{MyTa1999,MyEtAl1999,StGeHe2000,GeLoHe2001}.

%
%

%
%
\section{The ``Epsilon Method''}
\label{MathModel}

In QED bound-state calculations, we are often faced with the following 
problem: how to separate terms of a given order in the 
$(Z\alpha)$-expansion, and how to treat infrared divergences that
originate from higher-order terms
in the expansion of the bound-electron propagator in powers of the binding
field.

The so-called ``$\epsilon$-method'' has
been employed
in the analytic calculation of self-energy effects in bound systems.
The energy scales for the 
virtual photons are treated separately: 
(i) the nonrelativistic domain, in which the 
virtual photon assumes values of the order of the atomic binding energy,
and (ii) the relativistic domain, in which the virtual photon assumes values
of the order of the electron rest mass. 
The two energy domains are separated via a parameter $\epsilon$.
Without appropriate approximations and
expansions applicable to 
the two energy domains, respectively,
the analytic evaluation of either the high- or the low-energy
part would not be feasible. 

In one-photon calculations, we have to deal with one virtual photon 
energy $\omega$. For two-photon problems~\cite{Pa2001,JePa2002,JeNa2002},
one has to generalize the method to the case of two virtual
quanta and, by consequence, two separation parameters $\epsilon_1$
and $\epsilon_2$ (see also Fig.~1 of~\cite{Pa2001}). 
In both the one- and the two-photon case, we require the dependence on
the separation parameters to vanish at the end of the calculation,
i.e.~after the high- and the low-energy parts are added.

We follow here the discussion in App.~A of Ref.~\cite{JePa2002},
and we consider a model problem with only one ``virtual photon''.
In contrast to Ref.~\cite{JePa2002},
we choose a mathematical model problem of a slightly
more complex structure in
order to illustrate the occurrence of double-logarithmic terms
in the {\em semi-analytic} expansion, which involves powers and 
logarithms of the expansion parameter. The mathematical model
example reads
\begin{equation}
\label{J}
J(\beta) = \int_0^1 \ln(\omega) \,
\sqrt{\frac{\omega^2 + \beta^2}{1 - \omega^2}} \, 
{\mathrm{d}}\omega\,,
\end{equation}
where the integration variable $\omega$ might be interpreted as the 
``energy'' of a ``virtual photon''. 

We intend
to derive a semi-analytic expansion of $J(\beta)$ in powers of 
$\beta$ and $\ln \beta$. The quantity $Z\alpha$, which parameterizes
the strength of the binding Coulomb field,
replaces the expansion parameter $\beta$ in actual self-energy calculations. 
The ``high-energy part'' of the calculation
is given by the expression
\begin{equation}
\label{JH}
J_{\mathrm{H}}(\beta) = 
\int_\epsilon^1 \ln(\omega) \, 
\sqrt{\frac{\omega^2 + \beta^2}{1 - \omega^2}} \,
{\mathrm{d}}\omega\,.
\end{equation}
In the integration domain $\omega \in (\epsilon,1)$,
we may expand
\begin{equation}
\label{expansion1}
\sqrt{\omega^2 + \beta^2} = \omega + \frac{\beta^2}{2\,\omega} +
  \frac{\beta^4}{8\,\omega^3} + \mathcal{O}(\beta^6)\,.
\end{equation}
However, this expansion is not applicable in higher
orders to the domain $\omega \in (0,\epsilon)$ because of the appearance
of inverse powers of $\omega$ which lead to
``infrared divergences''.

We expand the integrand of $J_{\mathrm{H}}(\beta)$
first in powers of $\beta$ according to Eq.~(\ref{expansion1}).
The resulting integrals can be evaluated
analytically. 
Every term in the $\beta$-expansion is then 
expanded in powers of $\epsilon$ up to
the order $\epsilon^0$. Higher-order terms in $\epsilon$ are irrelevant;
they cancel at the end of the calculation (just as the divergent
terms in $\epsilon$), because the original expression
for the integral $J(\beta)$ in Eq.~(\ref{J})
is manifestly independent of $\epsilon$.
The result of the calculation of $J_{\mathrm{H}}(\beta)$ is
\begin{eqnarray}
\label{JHresult}
J_{\mathrm{H}}(\beta,\epsilon) &=& \left\{ \ln(2) - 1 
+ {\mathcal{O}}(\epsilon^2 \, \ln\epsilon) \right\} +
\beta^2 \, \left\{ - \frac14\, \ln^2(\epsilon)
+ \frac14\, \ln^2(2)
- \frac{\pi^2}{48} 
+ {\mathcal{O}}(\epsilon^2 \, \ln\epsilon) \right\}  
\nonumber\\[2ex]
& & + \beta^4 \, \left\{ \frac{1}{32} \, \ln^2\left( \epsilon \right)
- \frac{1}{16 \, \epsilon^2} \, \ln\left( \epsilon \right)
- \frac{1}{32 \epsilon^2}
- \frac{1}{32} \, \ln^2\left( 2 \right)
\right. \nonumber\\[1ex]
& & \quad \qquad \left.
+ \frac{1}{32} \, \ln\left( 2 \right)
+ \frac{\pi^2}{384} + \frac{1}{64}
+ {\mathcal{O}}(\epsilon^2 \, \ln \epsilon) \right\}
+ {\mathcal{O}}(\beta^6 \, \ln^2 \beta) \,.
\end{eqnarray}
The contribution $J_{\mathrm{H}}(\beta)$ corresponds to the 
``high-energy part'' in analytic self-energy calculations, where
the propagator of the bound electron may be expanded in
powers of $Z\alpha$ [see Sec.~III of~\cite{JeSoMo1997}].

The expression for the 
low-energy part $\omega \in (0,\epsilon$) reads
\begin{equation}
\label{JL}
J_{\mathrm{L}}(\beta) = 
\int_0^\epsilon \ln(\omega) \,
\sqrt{\frac{\omega^2 + \beta^2}{1 - \omega^2}} \,
{\mathrm{d}}\omega\,.
\end{equation}
We have to keep the numerator of the integrand
$\sqrt{\omega^2 + \beta^2}$ in unexpanded form. However, 
within the integration domain $\omega \in (0, \epsilon)$,
we may expand
the denominator $\sqrt{1 - \omega^2}$ of the integrand in powers
of $\omega$. Because $\omega < \epsilon$,
the expansion in powers
of $\omega$ is in fact an expansion in $\beta$ in 
the low-energy part.

One may draw an analogy between
the term $\sqrt{\omega^2 + \beta^2}$ and the 
Schr\"{o}dinger--Coulomb propagator in self-energy 
calculations~\cite{Pa1993,JePa1996,JeSoMo1997,JePa2002}.
In the low-energy domain, this propagator may {\em not} be expanded in powers
of the binding field. The expansion
\begin{equation}
\label{expansion2}
\frac{1}{\sqrt{1 - \omega^2}} = 1 + \frac{\omega^2}{2} + 
\frac{3}{8}\,\omega^4 + {\mathcal{O}}(\omega^6)
\end{equation}
corresponds to the $(Z\alpha)$-expansion
in the low-energy part. In actual self-energy
calculations (a detailed discussion can be found in Ref.~\cite{JeSoMo1997}), 
higher-order terms in the low-energy part
originate from the generalized 
Foldy--Wouthuysen transformation of the transition current,
from relativistic corrections to the Foldy--Wouthuysen transformed
Hamiltonian, 
and higher-order terms in the multipole expansion.
Specifically, more details concerning the
multipole expansion can be found
in the discussion following Eq.~(11) of Ref.~\cite{JeSoMo1997}.

We obtain for the low-energy part,
\begin{eqnarray}
\label{JLresult}
J_{\mathrm{L}}(\beta,\epsilon) &=& 
\beta^2 \, 
\left\{ \frac{1}{4} \, \ln^2\left( \epsilon \right)
- \frac{1}{4} \, \ln^2\left( \frac2\beta \right) 
- \frac{1}{4} \, \ln\left( \frac2\beta \right) 
- \frac{1}{8}
- \frac{\pi^2}{24} 
+ {\mathcal{O}}(\epsilon^2 \, \ln\epsilon) \right\}
\nonumber\\[2ex]
& & 
+ \beta^4 \, \left\{ - \frac{1}{32} \, \ln^2\left( \epsilon \right)
+ \frac{1}{16 \, \epsilon^2} \, \ln\left( \epsilon \right)
+ \frac{1}{32 \, \epsilon^2}
+ \frac{1}{32} \, \ln^2\left( \frac2\beta \right) \right.
\nonumber\\[1ex]
& & \quad \qquad \left.
- \frac{1}{64} \, \ln\left( \frac2\beta \right)
+ \frac{\pi^2}{192} - \frac{5}{256}
+ {\mathcal{O}}(\epsilon^2 \, \ln\epsilon) \right\}
+ {\mathcal{O}}(\beta^6 \, \ln^2 \beta) \,.
\end{eqnarray}
When the high-energy part (\ref{JHresult}) and the low-energy part 
(\ref{JLresult}) are added, the dependence on $\epsilon$ cancels,
and we obtain the result 
\begin{eqnarray}
\label{JResult}
J(\beta) &=& J_{\mathrm{H}}(\beta, \epsilon) + 
J_{\mathrm{L}}(\beta, \epsilon) \nonumber\\[2ex]
&=& \ln(2) - 1 \nonumber\\[2ex]
& & + \beta^2 \, \left\{ - \frac{1}{4}\, \ln^2\left(\frac{2}{\beta}\right) 
- \frac{1}{4} \, \ln\left(\frac{2}{\beta}\right)
+ \frac{1}{4} \, \ln^2\left(2\right)  
- \frac{\pi^2}{16}
- \frac{1}{8} \right\}  
\nonumber\\[2ex]
& & + \beta^4 \, \left\{ \frac{1}{32}\, \ln^2\left(\frac{2}{\beta}\right)
- \frac{1}{64} \, \ln\left(\frac{2}{\beta}\right)
- \frac{1}{32} \, \ln^2\left(2\right)  
+ \frac{1}{32} \, \ln\left(2\right) \right.
\nonumber\\[1ex]
& & \quad \qquad \left. - \frac{\pi^2}{128}
- \frac{1}{256} \right\} + 
{\mathcal{O}}(\beta^6 \, \ln^2 \beta)\,.
\end{eqnarray}
This result clearly demonstrates the semi-analytic character of the 
expansion: it involves double-logarithmic terms 
$\ln^2\left(2/\beta\right)$ and single logarithms 
$\ln\left(2/\beta\right)$ as well as constant terms.
The same pattern is observed in actual self-energy calculations.

%
%
\section{One-- and Two--Loop Self--Energy Calculations}
\label{SelfEnergy}

It is well known that the one-photon self-energy correction to the 
binding energy in low-$Z$ hydrogenlike systems can be parameterized as
as
\begin{equation}
\label{OnePhotonESE}
\delta E^{\rm (1\gamma)}_{\rm SE} = 
\frac{\alpha}{\pi} \, \frac{(Z \alpha)^4 \, m}{n^3} \, 
F(Z\alpha)\,,
\end{equation}
where the dimensionless quantity
$F(Z\alpha)$ has the following semi-analytic expansion,
\begin{eqnarray}
\label{OnePhotonF}
F(Z\alpha) &=& A_{41} \, \ln(Z \alpha)^{-2} + A_{40} +
  (Z \alpha) \, A_{50} \nonumber \\[2ex]
& &  + (Z \alpha)^2 \left[A_{62} \, \ln^2(Z \alpha)^{-2} +
A_{61} \,\ln(Z \alpha)^{-2} + A_{60} +
{\cal R} \right]\,,
\end{eqnarray} 
where ${\cal R}$ vanishes as $Z\alpha \to 0$.
The $A$-coefficients are state-dependent. In the following,
we focus on P states. The coefficients
$A_{41}$ and $A_{62}$ vanish for P states and states with 
higher orbital angular momenta.

In order to illustrate the analogy of our mathematical
model example (see Sec.~\ref{MathModel})
with one-loop self-energy calculations,
we give here the high-and low-energy parts derived
in~\cite{JePa1996} for the self-energy of an electron
bound in a hydrogenlike system
(the $2{\mathrm{P}}_{1/2}$ state):
\begin{equation}
F_{\mathrm{H}}(2{\mathrm{P}}_{1/2}) 
= -\frac{1}{6} + (Z \alpha)^2
\left[- \frac{2}{9 \, \epsilon}  -
\frac{103}{180} \ln{(\epsilon)} +
\frac{4177}{21600} - \frac{103}{180} \ln(2)
+ {\mathcal{O}}(\epsilon) \right] + {\mathcal{O}}(Z \alpha)^3\,,
\end{equation}
and
\begin{equation}
F_{\mathrm{L}}(2{\mathrm{P}}_{1/2})
= -\frac{4}{3} \, \ln k_0 (2{\mathrm{P}}) +
\left(Z \alpha \right)^2 \left[ \frac{2}{9 \, \epsilon}  +
\frac{103}{180} \ln\left( \frac{\epsilon}{(Z \alpha)^{2}} \right) 
-0.79569(1) +
{\mathcal{O}}(\epsilon) \right] + {\mathcal{O}}(Z \alpha)^3\,.
\end{equation}
Adding the two contributions, the dependence on $\epsilon$ cancels,
just as we had observed when forming the sum of the high-energy part
(\ref{JHresult}) and the low-energy result (\ref{JLresult}).

The two-loop self-energy correction to the energy of hydrogenlike
systems with low nuclear charge number reads
\begin{equation}
\label{MathModelDeltaE}
\delta E^{\rm (2\gamma)}_{\rm SE} = 
\left(\frac{\alpha}{\pi}\right)^2 \,
(Z\alpha)^4 \, \frac{m}{n^3} \, H(Z\alpha)\,,
\end{equation}
where the dimensionless function $H(Z\alpha)$ is given by
\begin{eqnarray}
\label{MathModelH}
H(Z\alpha) &=& B_{40}
\nonumber\\[2ex]
& & + (Z\alpha)^2 \, \left[ B_{63} \, \ln^3(Z\alpha)^{-2} + 
B_{62} \, \ln^2(Z\alpha)^{-2} +
B_{61} \, \ln(Z\alpha)^{-2} + B_{60} +
{\cal R}' \right]\,,
\end{eqnarray}
where ${\cal R}'$ vanishes as $Z\alpha \to 0$.
In two-photon calculations, we introduce two separation parameters
$\epsilon_1$ and $\epsilon_2$. This leads to four different
integration regions: (i) both photon energies are small 
(the ``low-and-low-energy part''), (ii)$+$(iii) two mixed
contributions (with one large and one small photon energy),
and (iv) a ``high-and-high-energy part'' with two 
large photon energies.

$B_{63}$ vanishes for all 
P, D, F, G, $\dots$ states, i.e.~for all atomic states with
a nonvanishing orbital angular momentum, and we first discuss here the the
coefficient $B_{62}$. The low-and-low-energy
part of the two-loop problem
(both virtual photons have a small energy) can be evaluated
using nonrelativistic quantum electrodynamics (NRQED). The relevant expression
is given in Eq.~(16) of Ref.~\cite{Pa2001}. The following double-logarithmic
term of order $\alpha^2\,(Z\alpha)^6$ originates from the low-energy part
[see Eq.~(38) of Ref.~\cite{JeNa2002}],
\begin{eqnarray}
{\mathcal L} = \left( \frac{\alpha}{\pi} \right)^2 \,
\ln\left[ \frac{\epsilon_1}{(Z\alpha)^2} \right] \,
\ln\left[ \frac{\epsilon_2}{(Z\alpha)^2} \right] \,
  \frac{2 \pi \, \langle \Delta \delta^{(3)}(\bbox{r}) \rangle}{9 \, m^4}\,,
\end{eqnarray}
where the known result for the matrix element reads
\begin{eqnarray}
\label{Laplace2}
\langle \Delta \delta^{(3)}(\bbox{r}) \rangle
\equiv 
\left. \Delta \left[\left| \phi_{n,l=1,m}(\bbox{r}) \right|^2 \right]
\right|_{r=0} = \frac{2}{3\pi}
\left[ (Z \alpha)^5 m^5 \right] \frac{n^2-1}{n^5}\,.
\end{eqnarray}
Because $B_{63}$ vanishes for P states, this result for the 
low-low-energy part determines uniquely the total result for 
$B_{62}$. This is because the dependence on $\epsilon_1$
and $\epsilon_2$ necessarily has to cancel at the end of the 
calculation according to 
\begin{eqnarray}
\label{simpl}
\lefteqn{\ln \left[\frac{\epsilon_1}{(Z\alpha)^2 m}\right] 
\ln \left[\frac{\epsilon_2}{(Z\alpha)^2 m}\right] +
\ln \left(\frac{m}{\epsilon_2}\right) \, 
\ln \left[\frac{\epsilon_1}{(Z\alpha)^2 m}\right]}
\nonumber\\[1ex]
& & + \ln \left(\frac{m}{\epsilon_1}\right) \, 
\ln \left[\frac{\epsilon_2}{(Z\alpha)^2 m}\right] +
\ln \left(\frac{m}{\epsilon_1}\right) \, 
\ln \left(\frac{m}{\epsilon_2}\right) = 
\ln^2 [(Z\alpha)^{-2}]\,.
\end{eqnarray}
Note that the logarithm $\ln[\epsilon_i/(Z\alpha)^2 m]$
is characteristic of the low-energy domain ($i=1,2$), whereas the 
logarithm $\ln(m/\epsilon_i)$
is characteristic of infrared divergencies that occur in the 
evaluation of integrals involving highly energetic virtual photons.
In view of (\ref{simpl}), we may conclude that the coefficient of 
\[
\ln\left[ \frac{\epsilon_1}{(Z\alpha)^2} \right] \,
\ln\left[ \frac{\epsilon_2}{(Z\alpha)^2} \right]  
\]
in the low-and-low-energy part is the same as the total
coefficient of $\ln^2 [(Z\alpha)^{-2}]$ for the two-loop
self energy. This has lead to a rigorous derivation of 
$B_{62}$ for P states~\cite{JeNa2002}, confirming the results of the
previous investigation~\cite{Ka1996},
\begin{equation}
\label{B62}
B_{62}(n{\rm P}) = \frac{4}{27} \, \frac{n^2-1}{n^2}\,.
\end{equation}
This result is valid for all P states 
independent of the electron spin.

The $\epsilon$-method provides a convenient tool for the
analysis of the problematic nonlogarithmic corrections
$A_{60}$ and $B_{60}$. Let us recall that the evaluation of
the one-loop coefficient $A_{60}$ for S states has attracted 
attention over many years~\cite{ErYe1965a,ErYe1965b,Er1971,Sa1981,Pa1993}. 
Today, we can hope to evaluate the corresponding $B_{60}$-coefficients
using this method. 

As a first step in this direction, 
we have obtained results~\cite{JePa2002} for the
following fine-structure differences of $B_{6k}$-coefficients
($k=0,1$) of P states,
\begin{eqnarray}
\Delta_{\mathrm{fs}} B_{61} &=&
  B_{61}(n {\mathrm P}_{3/2}) - B_{61}(n {\mathrm P}_{1/2})\,,
\nonumber\\[1ex]
\Delta_{\mathrm{fs}} B_{60} &=&
  B_{60}(n {\mathrm P}_{3/2}) - B_{60}(n {\mathrm P}_{1/2})\,.
\end{eqnarray}
We implicitly define the symbol $\Delta_{\mathrm{fs}}$ to 
denote the difference of the coefficient for an $n{\rm P}_{3/2}$
state and the corresponding $n{\rm P}_{1/2}$ level.
Note that the fine-structure difference of the leading
double logarithm vanishes [see Eq.~(\ref{B62})],
\begin{equation}
\Delta_{\rm fs} B_{62} = 0\,.
\end{equation}
Certain simplifications are possible when
considering the fine-structure
difference of the $B_{61}$ and $B_{60}$-coefficients. 
Specifically, the contribution of the 
high-and-high-energy integration domain can be
investigated with the help of a 
modified Dirac Hamiltonian. In this context, vertex corrections
are taken into account by considering the electron
form factor $F_1$ and $F_2$. A further simplification 
occurs because it is possible to devise a unified treatment
for both the low-and-low-energy domain and the mixed-energy
contributions. Some of the mixed-energy 
effects can be described by magnetic
form-factor corrections to the leading one-photon self-energy.
Because the magnetic form factor $F_2$ does not have infrared
divergences, contributions of the type
\[
\ln \left(\frac{m}{\epsilon_i}\right) \,
\ln \left[\frac{\epsilon_{3-i}}{(Z\alpha)^2 m}\right]
\qquad \qquad (i=1,2)
\]
vanish for the fine-structure difference.
We are therefore left with an infrared-safe and (in the context of the 
effective form-factor treatment) also ultraviolet-safe
mixed-energy integration domain, for which a simplified treatment
is possible.

We recall here the relevant results from~\cite{JePa2002}.
The high-and-high-energy integration domain yields
\begin{equation}
\label{EH}
E_{\mathrm H} = E_1 + E_{2a} + E_{2b} + E_3\,,
\end{equation}
where the correction $E_1$ is due to the $F_1$ form factor, 
$E_{2a}$ and $E_{2b}$ are due to the magnetic $F_2$ form factor,
and $E_3$ is caused by a second-order effect involving two one-loop  
magnetic form factor corrections to the spin-orbit interaction.
The results read~\cite[Eq.~(23)]{JePa2002}
\begin{equation}
\label{E1}
E_1 = \left(\frac{\alpha}{\pi}\right)^2 \,
  \frac{(Z\alpha)^6}{n^3} \, \left[ - F_1'^{(4)}(0) \,
     \frac{n^2 - 1}{n^2} \right] \, m^3\,,
\end{equation}
and according to~\cite[Eq.~(27)]{JePa2002}
\begin{equation}
\label{E2a}
E_{2a} = \left(\frac{\alpha}{\pi}\right)^2 \,
  \frac{(Z\alpha)^6}{n^3} \, \left[ F_2^{(4)}(0) \,
     \left( \frac{487}{720} + \frac{5}{8 n} -
        \frac{23}{20 n^2} \right) \right]\,m\,,
\end{equation}
as well as~\cite[Eq.~(36)]{JePa2002}
\begin{equation}
\label{E2b}
E_{2b} = \left(\frac{\alpha}{\pi}\right)^2 \,
  \frac{(Z\alpha)^6}{n^3} \, \left[
     - \frac{1}{6}\,\frac{n^2-1}{n^2}\,
\left( \ln\frac{m}{2\epsilon_1} + \ln\frac{m}{2\epsilon_2}\right) -
          \left(\frac{5}{18} + 2\,{\mathcal F}_2'^{(4)}(0)\,m^2 \right) \,
    \frac{n^2-1}{n^2} \right]\,m\,,
\end{equation}
and according to~\cite[Eq.~(30)]{JePa2002}
\begin{equation}
\label{E3}
E_3 = \left( \frac{\alpha}{\pi}\right)^2 \, \frac{(Z\alpha)^6}{n^3} \,
  \left[ \frac{227}{2880} + \frac{1}{32 n} - \frac{3}{80 n^2} \right]\,m\,.
\end{equation}
Analytic results are known~\cite{ReVa2000,GeRe2001,MaRe2001}
for the two-loop form factors entering into these expressions
(we take into account only the diagrams in Fig.~\ref{fig1} and
ignore the vacuum-polarization insertion in the 
one-loop vertex correction),
\begin{eqnarray}
\label{seF1p0}
m^2 \, F_1'^{(4)}(0) &=&
  - \frac{47}{576} - \frac{175}{144}\,\zeta(2)
  + 3\,\zeta(2) \, \ln 2 - \frac{3}{4} \, \zeta(3)\,, \\[3ex]
\label{seF20}
F_2^{(4)}(0) &=&
  - \frac{31}{16} + \frac{5}{2}\,\zeta(2)
  - 3\,\zeta(2) \, \ln 2 + \frac{3}{4} \, \zeta(3)\,, \\[3ex]
\label{seF2p0}
m^2 \, {\cal F}_2'^{(4)}(0)  &=&
    - \frac{151}{240} + \frac{61}{40}\,\zeta(2)
    - \frac{23}{10} \, \zeta(2) \, \ln 2
    + \frac{23}{40} \, \zeta(3)\,.
\end{eqnarray}
The sum of the low-and-low-energy domain and the 
mixed integration regions is 
\begin{equation}
\label{EL}
E_{\mathrm L} = E_4 + E_5\,,
\end{equation}
where the contribution $E_4$ reads~\cite[Eq.~(51)]{JePa2002}
\begin{equation}
\label{E4}
E_4 = E_{4a} + E_{4b} = \left( \frac{\alpha}{\pi} \right)^2 \,
\frac{(Z\alpha)^6 \, m}{n^3} \, \left[ - \frac{n^2 - 1}{6 n^2} \,
\left( \ln \frac{2 \epsilon_1}{(Z\alpha)^2\,m} +
\ln \frac{2 \epsilon_2}{(Z\alpha)^2\,m} \right) 
     + \frac{n^2 - 1}{n^2} \, \Delta_{\mathrm{fs}}{{\ell}}_4(n) \right]\,,
\end{equation}
and the explicit results for the $\ell_4(n)$ 
are given by~\cite{JeSoMo1997,JeLBMoSoIn2001}
\begin{eqnarray}
\Delta_{\mathrm{fs}}{\ell}_4(2) &=& 0.512~559~769(1)\,, \nonumber\\
\Delta_{\mathrm{fs}}{{\ell}}_4(3) &=& 0.513~111~333(1)\,, \nonumber\\
\Delta_{\mathrm{fs}}{{\ell}}_4(4) &=& 0.516~095~539(1)\,, \nonumber\\
\Delta_{\mathrm{fs}}{{\ell}}_4(5) &=& 0.518~940~860(1)\,.
\end{eqnarray}
$E_5$ reads~\cite[Eq.~(56)]{JePa2002}
\begin{equation}
\label{E5}
E_5 = \left( \frac{\alpha}{\pi} \right)^2 \,
\frac{(Z\alpha)^6 \, m}{n^3} \, \left[
\frac{n^2 - 1}{n^2} \, \Delta_{\mathrm{fs}}{{\ell}}_5(n) \right]\,,
\end{equation}
where
\begin{eqnarray}
\Delta_{\mathrm{fs}}{{\ell}}_5(2) &=& -0.173~344~868(1)\,, \nonumber\\
\Delta_{\mathrm{fs}}{{\ell}}_5(3) &=& -0.164~776~514(1)\,, \nonumber\\
\Delta_{\mathrm{fs}}{{\ell}}_5(4) &=& -0.162~263~216(1)\,, \nonumber\\
\Delta_{\mathrm{fs}}{{\ell}}_5(5) &=& -0.161~165~602(1)\,.
\end{eqnarray}
Adding all contributions $E_1$ -- $E_5$, the dependence on both
$\epsilon_1$ and $\epsilon_2$ cancels, and we obtain
\begin{equation}
\Delta_{\mathrm{fs}} B_{\mathrm 61} =
  - \frac{n^2 - 1}{3 n^2} 
\end{equation}
as well as
\begin{eqnarray}
\label{final}
\lefteqn{\Delta_{\mathrm{fs}} B_{\mathrm 60} =
   \left( \frac{227}{2880} + \frac{1}{32 n} - \frac{3}{80 n^2} \right) +
   F_2^{(4),S}(0) \, \left( \frac{487}{720} + \frac{5}{8 n} -
        \frac{23}{20 n^2} \right)}
\nonumber\\[1ex]
& &  + \frac{n^2 - 1}{n^2} \,
       \left[- \left( F_1'^{(4),S}(0) +
            2\, {\mathcal F}_2'^{(4),S}(0) \right) \, m^2
          - \frac{5}{18} + \Delta_{\mathrm{fs}}{{\ell}}_4(n) +
            \Delta_{\mathrm{fs}}{{\ell}}_5(n) \right]\,.
\end{eqnarray}
The explicit results for the principal
quantum numbers $n=2$--5 read
\begin{eqnarray}
\Delta_{\mathrm{fs}} B_{\mathrm 60}(2) &=& -0.361~196~470(1)\,,\\
\Delta_{\mathrm{fs}} B_{\mathrm 60}(3) &=& -0.410~149~385(1)\,,\\
\Delta_{\mathrm{fs}} B_{\mathrm 60}(4) &=& -0.419~926~624(1)\,,\\
\Delta_{\mathrm{fs}} B_{\mathrm 60}(5) &=& -0.420~872~513(1)\,.
\end{eqnarray}
These results have recently been generalized to the case of 
helium~\cite{PaSa2002} and used for an estimate of higher-order
binding corrections to the large and small fine-structure intervals
of the triplet P levels.

At this point we would like to mention
the further recent progress in the understanding of 
higher-order binding two-loop self-energy corrections 
(see Refs.~\cite{Ye2000,Ye2001,
YeSh2001}), which is faced with a number of conceptual 
and calculational difficulties, both in analytic
as well as numerical approaches.

%
%
\section{Squared Decay Rates}
\label{SquaredDecay}

An intriguing problem of bound-state quantum electrodynamics
is the interpretation of squared decay rates which follow from the 
Gell--Mann--Low--Sucher theorem~\cite{GMLo1951,Su1957} 
when applied to excited atomic states in two-loop order. 
We have recently shown (Ref.~\cite{JeEvKePa2002}) 
that the squared decay rates
cannot be interpreted in a natural way as real energy shifts.
Roughly speaking, the problems
in the interpretation originate from the fact that the 
Gell--Mann--Low--Sucher formalism involves {\em a priori}
asymptotic states with an infinite lifetime (vanishing decay
rate). The decay rate which enters in one-loop order
adds an imaginary part to the energy whose
square cannot be interpreted consistently as an energy shift 
within a formalism whose starting point was a theory that involves
asymptotic states with zero decay width (for a more detailed discussion
see Ref.~\cite{JeEvKePa2002})
Rather, a part of the problematic
corrections can be incorporated in a natural way in a modified
bound-electron Green function according to Eq.~(27) 
of~Ref.~\cite{JeEvKePa2002} which involves a ``decay-rate operator''
$\hat\Gamma$ defined in Eq.~(24) of Ref.~\cite{JeEvKePa2002}. 
Our formula (27) of Ref.~\cite{JeEvKePa2002}, which is equivalent
to Eq.~(16) on p.~218 of~Ref.~\cite{CTDRGr1992}, could be interpreted
to suggest that the modified Green function simply has a
pole on the second (unphysical) sheet of the Riemann surface
(see also the discussion on p.~217 of~Ref.~\cite{CTDRGr1992}). 
However, this is not the case: the only correction of the 
``squared-decay rate'' type which can be incorporated in a natural 
way into the electron Green function formalism is the one caused by
the loop-after-loop diagram in Fig.~\ref{fig1} (c), which is discussed
in Eqs.~(9) -- (14) of Ref.~\cite{JeEvKePa2002}. The interpretation
of the other problematic corrections of the type of a ``squared-decay rate'' 
discussed in Ref.~\cite{JeEvKePa2002} [see Eqs.~(5), (16),
and (18) {\em ibid.}] 
cannot be given as easily. In order to go beyond the predictive
limit of the current theory
set by the squared decay, one has to consider, in a fully gauge-invariant  
treatment, the excitation of the atom from the ground state
via the absorption of (laser) photons, and the return to the
ground state via spontaneous emission. Resummations of sets
of diagrams near resonance may be required. It has been stressed in
Ref.~\cite{JeEvKePa2002} that the ground state is the only
``true'' asymptotic state which may be used as an in- and out-state
in scattering theory. 

We would like to stress here that the two-time Green function 
method~\cite{Sh2002} avoids a number of problems associated with the
Gell--Mann--Low--Sucher theorem. Aside from the simplified
treatment of degenerate states, we would like to mention
the well-known fact that the infinitesimally 
damped $S_{\epsilon,\lambda}$-matrix
[see Eq.~(2) of~Ref.~\cite{JeEvKePa2002}] is, strictly speaking,
not renormalizable because the damping parameter $\epsilon$
breaks the covariance. 

A further problematic aspect of spectroscopic measurements is given
by the nonresonant corrections~\cite{Lo1952,LaSoPlSo2001,
JeMo2002,LaSoPlSo2002}. It has been stressed in~\cite{JeMo2002} that
nonresonant terms are enhanced in differential vs.~total cross sections,
and estimates for the effect in hydrogenic S--P transitions have
been obtained (see Sec.~3 of~\cite{JeMo2002}).
The enhancement of nonresonant effects in 
differential as compared to total cross sections
also follows in a natural way from the two-time Green function
formalism [see Eqs. (3.1.120) and (3.1.121) of~\cite{Sh2002}].
We observe that the experimental accuracy is approaching the 
1~MHz level at which the nonresonant terms become 
relevant (e.g.~\cite{EiWaHa2001}).

The order-of-magnitude at which the nonresonant terms enter in 
two-photon transitions depends crucially on the process under
study~\cite{JeMo2002,LaSoPlSo2002}. After the two-photon absorption,
the atom may return to the ground state via {\em spontaneous} emission of
two photons. In this case, the contribution of nonresonant terms
(on the level of $10^{-14}\,{\rm Hz}$)
is negligible at current and projected levels of 
experiment accuracy~\cite{JeMo2002}.
However, this does {\em not} imply that the experimental line shape
should remain Lorentzian up to this level of accuracy.
For this process, the dominant correction to the Lorentzian line-shape
in the two-photon transition
is given by radiative (not off-resonant!) corrections, and an estimate
of a relative contribution of order $\alpha \, (Z\alpha)^2$
has been given in~\cite{JeMo2002} (see Fig.~3 {\em ibid.},
this translates into $\sim 10^{-6}\,{\rm Hz}$ for the
1S--2S transition in atomic hydrogen).
In the current experiment~\cite{NiEtAl2000}, 
the excited hydrogen atom (2S) is quenched
to the rapidly decaying 2P state. This leads to an experimental
line width of the order of a kHz. In this case, the nonresonant
terms are enhanced, as argued in~\cite{LaSoPlSo2002}, and
an estimate of nonresonant corrections of the order of
$10^{-2}\,{\rm Hz}$ has been given for this {\em different}
experimental setup.

In all cases where off-resonant effects were considered, a formula of the 
general structure
\begin{equation} 
\label{oom}
\frac{[\mbox{experimental decay width} \, \Gamma]^2}
{\mbox{[typical atomic energy level difference} \, \Delta E]}
\end{equation}
has been obtained for the magnitude of the problematic shift
of the peak of the photon scattering cross section (which is a nonresonant
correction to the Lorentzian line shape). The meaning of the 
``typical atomic energy level difference'' depends on the process
under study: for differential cross sections in hydrogenic S--P 
transitions, a fine-structure level difference should be used for $\Delta E$
(see Ref.~\cite{JeMo2002}), whereas for 
total cross sections, the correct estimate is obtained by inserting
energy differences between states with a different principal quantum 
number (see Ref.~\cite{LaSoPlSo2001}). 
The order-of-magnitude estimate (\ref{oom}) implies that
(i) nonresonant effects are smaller than the experimental line
width by roughly a factor of $\Gamma/\Delta E$ and
(ii) the magnitude of the nonresonant terms decreases at
decreasing experimental line width. A formula
similar to (\ref{oom}) can be used in order to estimate the
order-of-magnitude of the energy level ``shift'' by squared 
decay rates.

%
%
\section{Conclusions}
\label{Conclusions}

We have discussed the evaluation of higher-order binding 
corrections to the one-and two-loop
self energy via the $\epsilon$-method (Sec.~\ref{MathModel})
This method has proven to be a useful calculational tool,
as it leads to a rather clear separation of terms
which contribute at different orders in the 
$(Z\alpha)$-expansion and to a transparent formulation
of the physical problem. A first step in a systematic
investigation of the highly problematic nonlogarithmic 
$B_{60}$-coefficient for hydrogenic bound states is presented.
The cancellation of the expansion parameter $\epsilon$ 
at the end of the calculation is demonstrated by way
of a mathematical model example [Eqs.~(\ref{J}) -- (\ref{JResult})]
and in concrete QED bound-state calculations at the 
one- and two-loop level (see Sec.~\ref{SelfEnergy}).

When performing two-loop self-energy calculations for excited
states, one is lead in a natural way to the problem of the 
interpretation of squared decay rates. These effects cannot be
interpreted self-consistently as radiative energy shifts of 
a specific atomic energy level (Sec.~\ref{SquaredDecay}).
We consider the connection to the Gell--Mann--Low--Sucher theorem
and to nonresonant corrections to the Lorentzian line shape.
The process dependence of corrections to the Lorentzian 
line shape in two-photon transitions is analyzed, and an
order-of-magnitude estimate for off-resonant corrections
to the peak of the photon scattering cross section is
given [see Eq.~(\ref{oom})].
 
%
%
\section*{Acknowledgements}

U.D.J.~and C.H.K.
gratefully acknowledge
funding by the Deutsche Forschungsgemeinschaft (Nachwuchsgruppe
within the Sonderforschungsbereich 276), and
K.P. acknowledges support from the
Polish Committee for Scientific Research under
contract no. 2P03B 057 18.

\end{document}